# H$_2$O line mapping at high spatial and spectral resolution

## *Herschel* observations of the VLA 1623 outflow *

P. Bjerkeli[1], R. Liseau[1], B. Larsson[2], G. Rydbeck[1], B. Nisini[3], M. Tafalla[4], S. Antoniucci[3], M. Benedettini[5], P. Bergman[6], S. Cabrit[7], T. Giannini[3], G. Melnick[8], D. Neufeld[9], G. Santangelo[3] and E. F. van Dishoeck[10,11]

[1] Department of Earth and Space Sciences, Chalmers University of Technology, Onsala Space Observatory, 439 92 Onsala, Sweden
[2] Department of Astronomy, Stockholm University, AlbaNova, 106 91 Stockholm, Sweden
[3] INAF - Osservatorio Astronomico di Roma, Via di Frascati 33, 00040 Monte Porzio Catone, Italy
[4] Observatorio Astronómico Nacional (IGN), Calle Alfonso XII,3. 28014, Madrid, Spain
[5] INAF - Instituto di Astrofisica e Planetologia Spaziali, Via Fosso del Cavaliere 100, 00133 Roma, Italy
[6] Onsala Space Observatory, Chalmers University of Technology, 439 92 Onsala, Sweden
[7] LERMA, Observatoire de Paris, UMR 8112 of the CNRS, 61 Av. de l'Observatoire, 75014 Paris, France
[8] Harvard-Smithsonian Center for Astrophysics, 60 Garden Street, Cambridge, MA 02138, USA
[9] Department of Physics and Astronomy, Johns Hopkins University, 3400 North Charles Street, Baltimore, MD 21218, USA
[10] Leiden Observatory, Leiden University, PO Box 9513, 2300 RA Leiden, The Netherlands
[11] Max Planck Institut für Extraterrestrische Physik, Gießenbachstraße 1, 85748 Garching, Germany

## ABSTRACT

*Context.* Apart from being an important coolant, water is known to be a tracer of high-velocity molecular gas. Recent models predict relatively high abundances behind interstellar shockwaves. The dynamical and physical conditions of the water emitting gas, however, are not fully understood yet. Using the *Herschel* Space Observatory, it is now possible to observe water emission from supersonic molecular outflows at high spectral and spatial resolution. Several molecular outflows from young stars are currently being observed as part of the WISH (Water In Star-forming regions with *Herschel*) key program.
*Aims.* We aim to determine the abundance and distribution of water, its kinematics, and the physical conditions of the gas responsible for the water emission. The observed line profile shapes help us understand the dynamics in molecular outflows.
*Methods.* We mapped the VLA 1623 outflow, in the ground-state transitions of $o$-H$_2$O, with the HIFI and PACS instruments. We also present observations of higher energy transitions of $o$-H$_2$O and $p$-H$_2$O obtained with HIFI and PACS towards selected outflow positions. From comparison with non-LTE radiative transfer calculations, we estimate the physical parameters of the water emitting regions.
*Results.* The observed water emission line profiles vary over the mapped area. Spectral features and components, tracing gas in different excitation conditions, allow us to constrain the density and temperature of the gas. The water emission originates in a region where temperatures are comparable to that of the warm H$_2$ gas ($T \gtrsim 200$ K). Thus, the water emission traces a gas component significantly warmer than the gas responsible for the low-$J$ CO emission. The water column densities at the CO peak positions are low, i.e. $N$(H$_2$O) $\simeq (0.03 - 10) \times 10^{14}$ cm$^{-2}$.
*Conclusions.* The water abundance with respect to H$_2$ in the extended outflow is estimated at $X$(H$_2$O) $< 1 \times 10^{-6}$, significantly lower than what would be expected from most recent shock models. The H$_2$O emission traces a gas component moving at relatively high velocity compared to the low-$J$ CO emitting gas. However, other dynamical quantities such as the momentum rate, energy, and mechanical luminosity are estimated to be the same, independent of the molecular tracer used, CO or H$_2$O.

## 1. Introduction

Outflows are spectacular signposts of the star formation process, where swept up material from the parental cloud, excited in the interaction with the ambient medium, can be observed on parsec scale distances from the central source (see e.g. Richer et al. 2000; Tafalla & Bachiller 2011, for reviews). Outflows from young stars are most likely a consequence of a re-distribution of angular momentum during the process of proto-stellar accretion. Jets launched from a region close to the proto-stellar disk

(Banerjee & Pudritz 2006) likely drives them and shock waves interacting with the ambient medium decelerate the gas in the aftermath of the shock. Emission from H$_2$O is produced in the post-shock medium where the molecules can cool on a relatively long time scale but the emission may also stem from within the jet itself (see e.g. Eislöffel et al. 2000). Momentum conservation implies that the bulk of the flow is moving at a moderate velocity and radial velocities should increase along the flow axis (see e.g. Cabrit et al. 1997). Previous spectrally resolved observations reveal complicated line profiles where gas moving at different velocities (with respect to the observer) show up as spectral features. The shape of the line profiles is a consequence of the geometry, abundance, and velocity fields but it also reflects

---

the optical depth in the observed lines. In some of the observed sources, high velocity components, hereafter referred to as bullets, show up (see e.g. Kristensen et al. 2011).

The water abundance in the outflow is expected to reflect the type of shock physics in action (see e.g. Bergin et al. 1998; Kaufman & Neufeld 1996; Hollenbach et al. 1989). Depending on the physical conditions in the flow, water abundances with respect to H$_2$ can vary by several orders of magnitude in the post-shock gas, and can be as high as $10^{-4}$ (Flower & Pineau des Forêts 2010). H$_2$O line profiles and abundances are often compared to those of CO and other shock tracers, readily observed with ground-based telescopes. However, recent observations with HIFI indicate that emission from low-$J$, CO and H$_2$O does not always probe gas with comparable excitation conditions. Likewise, emission from these two molecules does not necessarily correlate spatially. Whilst the CO emission plausibly traces entrained gas, it has been found that H$_2$ can be a valuable probe, tracing shocked gas at excitation conditions similar to that of H$_2$O (Santangelo et al. 2012; Nisini et al. 2010).

The radiation from most water transitions is effectively blocked out by the Earth's atmosphere so these transitions cannot be observed using ground-based facilities. Instead, space based observatories such as SWAS (Melnick et al. 2000) and Odin (Nordh et al. 2003) have been used to observe the lowest rotational transitions of water. Using these facilities, the abundance of this molecule has been constrained (Franklin et al. 2008; Bjerkeli et al. 2009). It has also been shown that the H$_2$O abundance is generally lower than what was previously expected.

In this paper we present a study of the outflow powered by VLA 1623 carried out with the *Herschel* Space Observatory (Pilbratt et al. 2010). VLA 1623 is located in the constellation of Ophiuchus at a distance of 120 pc (e.g. Lombardi et al. 2008) near the $\rho$ Oph A cloud core (Loren et al. 1990). This core is part of the star-forming cloud L1688, where a large number of stars have recently been formed (see e.g. Bontemps et al. 2001; Duchêne et al. 2004). Apart from being located in one of the most nearby star-forming regions it is also considered to be the prototype of a very young stellar object, i.e. a Class 0 source (André et al. 1990, 1993). The central source drives a collimated bipolar CO outflow extending more than 0.5 pc in the southeastern direction (André et al. 1990, 1993; Dent et al. 1995). The nature of the source itself has in recent years been a matter of debate. For instance, Looney et al. (2000) suggested that it is a binary system with two circumstellar disks. This has, however, been challenged by the interferometric observations presented in Ward-Thompson et al. (2011). An alternative interpretation is that the observed continuum sources close to VLA 1623 are knots in the north-western jet (Maury & André 2012).

Here we present HIFI (Heterodyne Instrument for the Far Infrared) and PACS (Photodetector Array Camera and Spectrometer) observations of the ground state transitions of *ortho*-water at 557 and 1670 GHz. Two maps were obtained, covering a $5' \times 2'$ region of the VLA 1623 outflow and the southern part of the $\rho$ Oph A cloud core. Also, a line survey was conducted towards two outflow positions. Combined, these datasets provide us with kinematic and spatial information at high resolution. This allows us to constrain the physical and dynamical parameters of the gas responsible for the observed H$_2$O emission.

The paper is organised as follows; the HIFI and PACS observations are presented in Section 2 and the main results are presented in Section 3. The obtained maps and line survey are discussed in detail in Section 4 where we also perform an excitation analysis towards selected positions in the outflow. In Secction 5, our main conclusions are briefly summarised.

## 2. Observations and data reduction

The mapping observations of VLA 1623 were obtained with the HIFI (de Graauw et al. 2010) and PACS (Poglitsch et al. 2010) instruments aboard *Herschel* between 17 February 2010 and 12 March 2011. In addition, selected transitions of *ortho*- and *para*-water were observed towards two outflow positions. These observations were carried out as part of the Water In Star-forming regions with Herschel (WISH) key program (van Dishoeck et al. 2011).

The data were processed using the Herschel Interactive Processing Environment (HIPE, Ott 2010). Baseline subtraction and subsequent data reduction were performed in XS[1] and Matlab.

### 2.1. HIFI

#### 2.1.1. HIFI - map

A map of the region surrounding VLA 1623 was obtained in the 557 GHz ground state rotational transition of $o$-H$_2$O on 1 September, 2010 (see Fig. 1). The 3.5 meter Cassegrain telescope has a Half Power Beam Width (HPBW) of 38″ at 557 GHz and a pointing accuracy of the order of 1″ (García-Lario et al. 2011). The calibration uncertainty is ∼10% (Roelfsema et al. 2012) and the main beam efficiency at 557 GHz has been measured (from observations of Mars) to 0.76 (Olberg 2010). The data were calibrated using the *Herschel Interactive Processing Environment* (HIPE) version 7.2.0. In this paper, data from the Wide Band Spectrometer (WBS) are used. The WBS is an acousto-optical spectrometer with a bandwidth coverage of 4 GHz. The spacing between the channels is 500 kHz which corresponds to 0.3 km s$^{-1}$ at 557 GHz. The HIFI map was obtained using the "On-The-Fly (OTF) Maps with Position-Switch Reference" observing mode, where the telescope is scanning the area. The reference spectrum was obtained toward a position which was located 14′ away. In Fig. 1, however, the data are presented on a rectangular grid. A Gaussian regridding procedure was used to compute the spectra at each position. The readout positions for the OTF map, used as input for the regridding, are presented in Fig. 2. Here, the positions for the horizontal (H) and vertical (V) positions were calculated separately.

The *Herschel* HPBW at 557 GHz is about a factor of 3 larger than that at 1670 GHz. To improve the spatial resolution, a Statistical Image Deconvolution (SID) method was used to deconvolve the HIFI map. This method is based on Bayes' theorem and is described in detail in Rydbeck (2008). The spatial resolution of the HIFI map, after de-convolution, is estimated at ∼20″ in the region where the signal-to-noise is high.

#### 2.1.2. HIFI - pointed observations

Selected transitions (see Table 1) were observed towards two outflow positions showing enhanced emisson in CO (2–1) (see André et al. 1990). These positions are referred to as Blue 1 (B1) and Red 2 (R2) hereafter and the offset with respect to VLA 1623 is (22″, –11″) for the B1 position and (–57″, +29″)

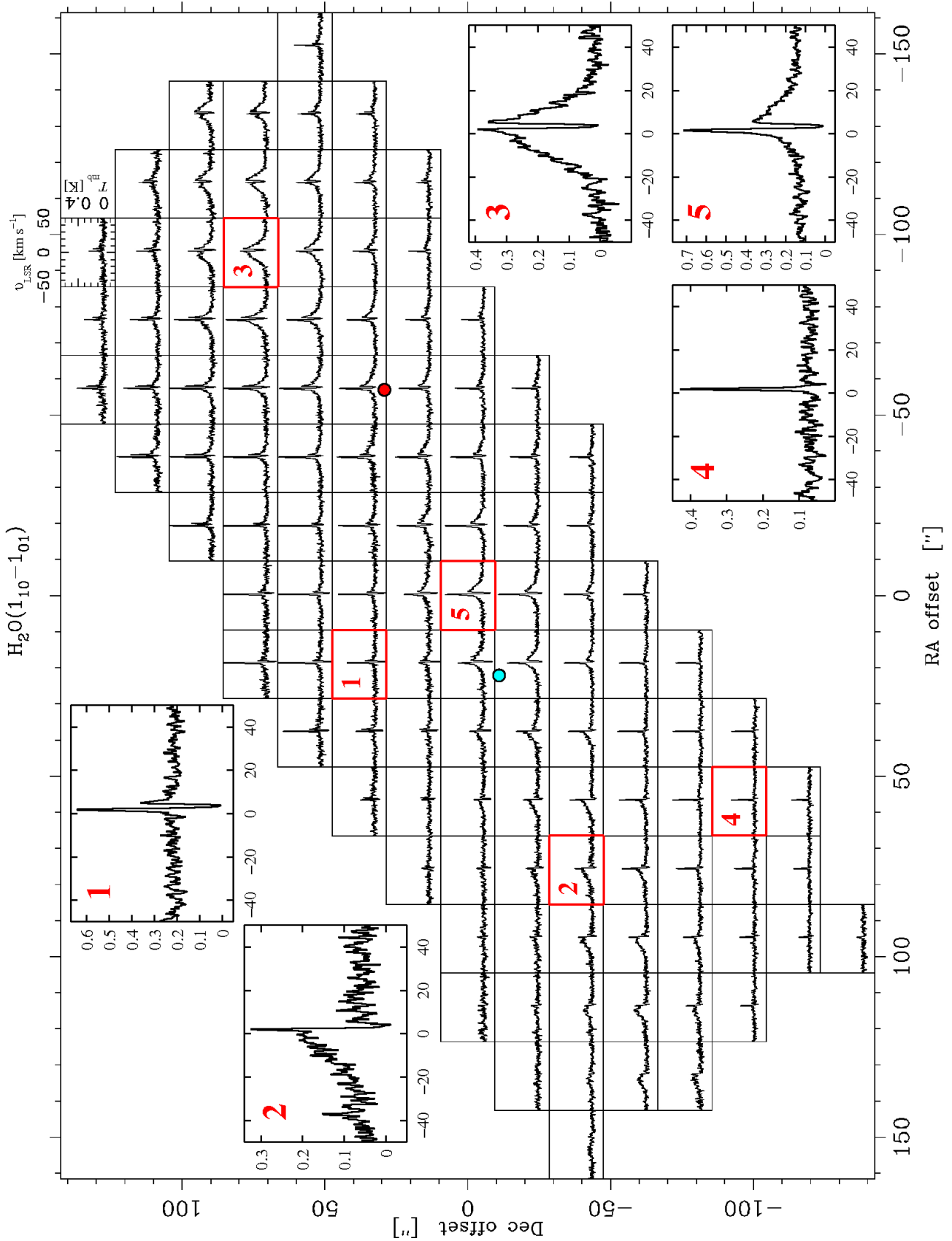

**Fig. 1.** H$_2$O $(1_{10} - 1_{01})$ map obtained with HIFI. The data have been regridded onto a regular grid in RA and Dec. The spacing between each position is 19″ and the map is centred on the Class 0 source VLA 1623 at $\alpha_{2000} = 16^h26^m26.4^s$, $\delta_{2000} = -24°24'31''$. All spectra are plotted on the $T_{mb}$-scale and velocities are with respect to the $v_{LSR}$ (from −50 to +50 km s$^{-1}$). Note that the spectra in the enlarged insets have not been baseline subtracted, but corrected such that the offset shows the single side-band continuum level. The positions of B1 and R2 are indicated with coloured dots.

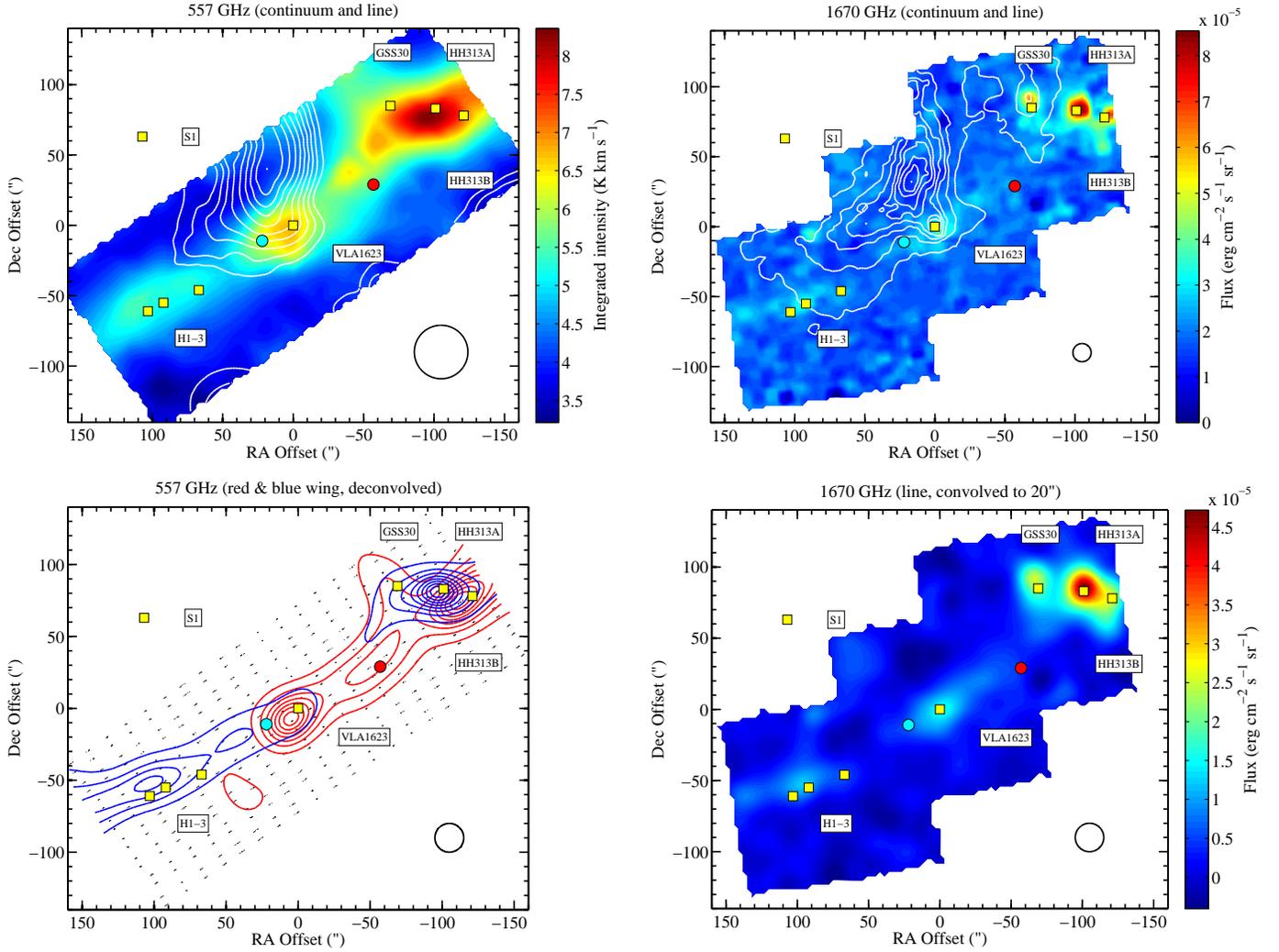

**Fig. 2.** *Upper left panel:* Colour map shows the integrated intensity over the H$_2$O ($1_{10} - 1_{01}$) line from $-40$ to $+55$ km s$^{-1}$. Contours show the level of the 557 GHz continuum in steps of 14 mK where the first contour is at 79 mK. The HPBW is 38″. *Upper right panel:* The colour map shows the flux of the H$_2$O ($2_{12} - 1_{01}$) line. Contours show the continuum level in steps of $4 \times 10^{-4}$ erg cm$^{-2}$ s$^{-1}$ $\mu$m$^{-1}$ sr$^{-1}$ where the first contour is at $4 \times 10^{-4}$ erg cm$^{-2}$ s$^{-1}$ $\mu$m$^{-1}$ sr$^{-1}$. The HPBW is 13″. *Lower left panel:* The blue and red contours show the integrated H$_2$O ($1_{10} - 1_{01}$) intensities in the blue and red wings for the de-convolved map. Red contours are in steps of 0.85 K km s$^{-1}$ and the first contour is at 0.16 K km s$^{-1}$. Blue contours are in steps of 1.14 K km s$^{-1}$ and the first contour is at 0.16 K km s$^{-1}$. The readout positions for the OTF map are indicated with black dots. The horizontal and vertical polarisations are separated by 6″ in the north-east to south-west direction. *Lower right panel:* The colour map shows the integrated intensity for the H$_2$O ($2_{12} - 1_{01}$) map convolved to 20″ resolution.

In all panels, the peak positions of the CO ($2-1$) emission are indicated with blue (B1) and red (R2) dots. The star S 1, the young stellar object GSS30 and other HH objects and near-infrared knots are indicated with labels and yellow squares. The beam sizes (or equivalent resolution for modified maps) are indicated with black circles in the lower right corners of each figure.

for the R2 position. The observations were carried out in "Dual Beam Switch (DBS)" mode where the beam is switched by the internal chopper to a sky position 3′ away from the target.

## 2.2. PACS

### 2.2.1. PACS - map

A PACS map of the 1670 GHz ground state transition of $o$-H$_2$O was obtained on 1 September, 2010 (see Fig. 2). The map was obtained in line spectroscopy mode and the wavelength range was centred on the H$_2$O ($2_{12} - 1_{01}$) transition at 179 $\mu$m. A map of the VLA 1623 outflow was obtained by combining three raster maps centred at three different positions along the axis of the outflow. Each map consists of 3×3 PACS frames with a spatial separation of 40″. The PACS instrument is a 5×5 integral field unit array having a spatial separation of 9″4 between each pixel. The H$_2$O ($2_{12} - 1_{01}$) line is not resolved at $R = 1700$ corresponding to a FWHM of the line of $\sim$175 km s$^{-1}$. The data were reduced using Hipe v.6.0. The observed fluxes were normalised to the telescope background and subsequently converted to absolute fluxes using Neptune as a calibrator. The uncertainty of the flux calibration is $\sim$30%. Software such as IDL and XS have been used to obtain continuum subtracted line maps.

### 2.2.2. PACS - pointed observations

Two *ortho*- and two *para*-transitions were observed using the PACS array in line spectroscopy mode (see Table 1) . These observations were carried out towards the blue- and red-shifted

**Table 1.** Observations carried out with PACS and HIFI.

| Instrument - Band | Transition | Spin isomer | Frequency (GHz) | $E_u/k_B$ (K) | HPBW/FOV$^a$ (″) | Date (YYMMDD) | $t_{int}$ (sec) | Date (YYMMDD) | $t_{int}$ (sec) |
|---|---|---|---|---|---|---|---|---|---|
| *Mapping observations* | | | | | | | | | |
| HIFI-1 | H$_2$O $(1_{10} - 1_{01})$ | *ortho* | 556.936 | 61.0 | 38 | 100901 | 6552 | | |
| PACS | H$_2$O $(2_{12} - 1_{01})$ | *ortho* | 1669.905 | 114.4 | 13 | 100227 | 11154 | | |
| *Single pointing* | | | | | | R2 (−57″,+29″) | | B1 (+22″,−11″) | |
| HIFI-1 | H$_2$O $(1_{10} - 1_{01})$ | *ortho* | 556.936 | 61.0 | 38 | 100929 | 878 | 100901 | 440 |
| HIFI-2 | H$_2$O $(2_{11} - 2_{02})$ | *para* | 752.033 | 136.9 | 28 | 110312 | 430 | 100916 | 454 |
| HIFI-4 | H$_2$O $(3_{12} - 3_{03})$ | *ortho* | 1097.365 | 249.4 | 19 | 110217 | 1658 | 110217 | 829 |
| HIFI-4 | H$_2$O $(1_{11} - 0_{00})$ | *para* | 1113.343 | 53.4 | 19 | 110217 | 765 | 110217 | 829 |
| HIFI-6 | H$_2$O $(2_{12} - 1_{01})$ | *ortho* | 1669.905 | 114.4 | 13 | 110218 | 1008 | 110218 | 1058 |
| PACS | H$_2$O $(3_{03} - 2_{12})$ | *ortho* | 1716.770 | 196.8 | 47 × 47 | 100227 | 3098 | 100226 | 3098 |
| PACS | H$_2$O $(3_{13} - 2_{02})$ | *para* | 2164.132 | 204.7 | 47 × 47 | 100227 | 3098 | 100226 | 3098 |
| PACS | H$_2$O $(4_{04} - 3_{13})$ | *para* | 2391.573 | 319.5 | 47 × 47 | 100227 | 3098 | 100226 | 3098 |
| PACS | H$_2$O $(2_{21} - 1_{10})$ | *ortho* | 2773.977 | 194.1 | 47 × 47 | 100227 | 3098 | 100226 | 3098 |

**Notes.** $^{(a)}$ For the PACS single pointing observations, the Field Of View (FOV) of the array is indicated in the table.

outflow lobes with the central spaxel centred on the B1 and R2 positions respectively. For the excitation analysis in Sec. 4.1, line fluxes have been extracted in a circular region (32″) centred on B1 and R2. Using this method, line fluxes should be accurate to within 30%.

## 3. Results

The H$_2$O $(1_{10} - 1_{01})$ and H$_2$O $(2_{12} - 1_{01})$ lines are detected towards the outflow emanating from VLA 1623 (Fig. 1 & 2). In addition, several other transitions of *o*-H$_2$O and *p*-H$_2$O are detected towards the CO peak positions, R2 and B1. We note, however, that these positions do not correspond to peaks in the H$_2$O emission. A summary of the observed H$_2$O lines is presented in Table 2.

### 3.1. HIFI map

The line profile shapes observed towards individual positions change significantly throughout the H$_2$O $(1_{10} - 1_{01})$ map obtained with HIFI. In Fig. 1, we present the profiles towards 114 positions with five positions of particular interest enlarged. In summary, the line profiles exhibit 3 different components possibly tracing gas that originate in different physical regions.

#### 3.1.1. Absorption feature

An absorption feature, which in some positions reaches below the observed continuum level, is visible across the map ($v_{LSR} \simeq 3.7$ km s$^{-1}$). The absorption is detected at the majority of the positions in the map and is most prominent in a region centred (+25″, +25″) offset from the position of VLA 1623, i.e. near SM1 (Motte et al. 1998). No clear, large-scale velocity differences for the absorption feature can be seen (see Fig. 3). The southern part of the core, where no or little outflowing gas is detected, is presented in spectrum 4 of Fig. 1 and shows hardly any absorption.

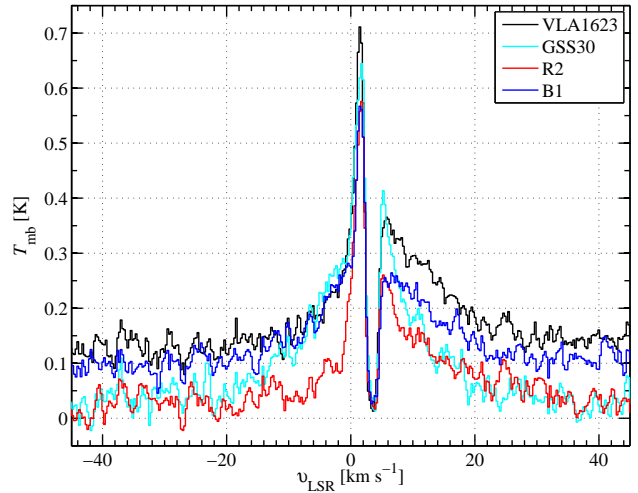

**Fig. 3.** H$_2$O $(1_{10} - 1_{01})$ spectra towards four selected positions in the map. No large-scale velocity gradients can be detected from the centroid velocity of the absorption feature. The absorption feature reaches below the continuum level where the continuum is strong, i.e. towards VLA 1623 and B1. Note that the spectra not have been baseline subtracted.

#### 3.1.2. Spectral wings

Towards the outflow, gas motions are clearly traced in the form of broad spectral wings. In the central region, both red-shifted and blue-shifted emission is detected. The blue-shifted part of the flow is free from contamination from other flows. Here, no red-shifted components are detected (see Fig. 1, spectrum 2). On the other hand, a blue-shifted component is clearly visible in the same region as the red part of the flow emanating from VLA 1623 (see Fig. 2 & spectrum 3 of Fig. 1). Here, the H$_2$O emission line profiles show a strong blue-red asymmetry. The blue-shifted component is most likely associated with GSS30 (see e.g. Tamura et al. 1990) and the peak of this emission is located close to the position of the Herbig-Haro object HH313 A (Davis & Eislöffel 1995). However, the presence of blue-shifted gas in the region close to R2 is not easily associated with GSS30.

**Table 2.** Integrated intensities of the lines observed with PACS and HIFI during the line survey. Numbers in parentheses are 1 $\sigma$ uncertainties.

| Transition | Frequency | $\Delta\upsilon_{LSR}$ | R2 | | B1 | |
|---|---|---|---|---|---|---|
| | | | $\int T_{mb}\,d\upsilon$ | $T_{mb,rms}{}^a$ | $\int T_{mb}\,d\upsilon$ | $T_{mb,rms}{}^a$ |
| | (GHz) | (km s$^{-1}$) | (K km s$^{-1}$) | (mK) | (K km s$^{-1}$) | (mK) |
| H$_2$O ($1_{10} - 1_{01}$) | 556.936 | $-30$ to 30 | 3.4 (0.03) | 8 | 3.6 (0.05) | 12 |
| H$_2$O ($2_{11} - 2_{02}$) | 752.033 | $-30$ to 30 | 0.3 (0.08) | 22 | 0.8 (0.08) | 21 |
| H$_2$O ($3_{12} - 3_{03}$) | 1097.365 | - | - | 28 | - | 30 |
| H$_2$O ($1_{11} - 0_{00}$) | 1113.343 | $-30$ to 30 | 1.4 (0.10) | 36 | 1.9 (0.10) | 36 |
| H$_2$O ($2_{12} - 1_{01}$) | 1669.905 | - | - | 106 | - | 106 |
| H$_2$O ($3_{03} - 2_{12}$) | 1716.770 | $-200$ to 200 | 0.7 | 0.60 | 0.6 | 1.05 |
| H$_2$O ($3_{13} - 2_{02}$) | 2164.132 | $-200$ to 200 | 0.2 | 0.19 | - | 0.31 |
| H$_2$O ($4_{04} - 3_{13}$) | 2391.573 | - | - | 0.19 | - | 0.22 |
| H$_2$O ($2_{21} - 1_{10}$) | 2773.977 | - | - | 0.29 | - | 0.15 |

**Notes.** $^{(a)}$ For the HIFI observations, the frequency bin size when calculating the rms, is the same as the channel spacing. For the PACS data, the frequency bin sizes at 1717, 2164, 2392, and 2774 GHz are 0.3, 0.5, 0.6, and 0.7 GHz respectively. The rms is in these cases measured in the central spaxel.

In fact, blue-shifted emission is detected at all positions along the red part of the outflow. This could indicate that the north-western flow is nearly in the plane of the sky (André et al. 1990). At $\upsilon_{LSR} < \pm 20$km s$^{-1}$, there is a trend where the velocity is increasing with distance from the central source (see Fig. 4).

### 3.1.3. High-velocity gas

Spectral features ("bullets") are detected towards several positions in the map (see e.g. spectrum 2 in Fig 1). These features trace gas at velocities higher than those traced by the outflow wings. There is, however, no tendency for the highest velocity gas to be observed at a certain distance from the central source. A channel map of the region reveals high velocity gas at positions far from as well as close to the driving source of the flow (see Fig. 4). It should be pointed out, that high velocity gas is also present in the vicinity of VLA 1623. This emission is slightly offset to the north and aligned with the axis of the outflow.

### 3.1.4. Narrow emission feature

The central emission feature ($\upsilon_{LSR} \approx 2$ km s$^{-1}$) is visible in the majority of positions in the observed map and diminishes along the blue flow axis. The velocity of the peak emission is lowest towards VLA 1623 ($\sim 1.6$ km s$^{-1}$). However, the absorption goes deep in this region and no firm conclusions can be drawn regarding the large-scale velocity field of this component. It is clearly affected both by the wing emission (at lower velocities) and the deep absorption (at higher velocities). The narrow emission feature is strong towards the central position (Fig. 1, spectrum 5) and peaks close to the H$_2$CO peak position (Bergman et al. 2011) slightly north from VLA 1623, i.e. at ($+17''$,$+76''$).

To the south of the blue-shifted outflow lobe, emission line profiles appear single peaked (Fig. 1, spectrum 4). A rough estimate of the water column density in this region can be obtained following the method used by Snell et al. (2000). Assuming effectively optically thin gas at a temperature of $T = 30$ K, we estimate the water column density in the ($+57''$,$-95''$) position to be $N(o$-H$_2$O$) = 1.3 \times 10^{20}/n$(H$_2$) cm$^{-2}$.

### 3.2. PACS map

The H$_2$O ($2_{12} - 1_{01}$) line emission also seems to be extended along the outflow (see Fig. 2). However, the highest signal-to-noise ratio is obtained towards the positions of GSS30, HH313 A and HH313 B. In addition, enhanced emission is detected close to the position of VLA 1623 (see Sec. 4.1.3).

### 3.3. H$_2$O line survey

#### 3.3.1. HIFI

Five H$_2$O transitions were observed with HIFI towards the B1 and R2 positions (see Fig. 5 and Table 2). The *ortho* H$_2$O ($1_{10} - 1_{01}$) line at 557 GHz ($E_u/k_B = 61$ K) is, as expected, detected at both positions with a high signal-to-noise ratio. The *para* H$_2$O ($1_{11} - 0_{00}$) line at 1113 GHz ($E_u/k_B = 53$ K) is also detected with a reasonable signal-to-noise ratio in both positions while the *para* H$_2$O ($2_{11} - 2_{02}$) line at 752 GHz ($E_u/k_B = 137$ K) has a rather low signal-to-noise ratio at both positions. At both positions we observe a difference in flux between the H and V polarisations for the latter two lines. The cause of this discrepancy is at present not understood, but could be due to the difference in pointing between the two polarisations and/or baseline issues. Consequently, we adopt a somewhat higher value on the uncertainty, viz. 40% for the 1113 GHz line and 50% for the 752 GHz line. The *ortho* H$_2$O ($2_{12} - 1_{01}$) line at 1670 GHz ($E_u/k_B = 114$ K) is not detected towards the R2 position but there is a possible detection towards B1 ($2\sigma$). The *ortho* H$_2$O ($3_{12} - 3_{03}$) line at 1097 GHz ($E_u/k_B = 249$ K) is not detected in any of the observed spectra.

#### 3.3.2. PACS

Four H$_2$O transitions were observed with PACS in two regions centred on B1 and R2 (see Fig. 6 and Table 2). Towards the B1 position, only the *ortho* H$_2$O ($3_{03} - 2_{12}$) line at 1717 GHz ($E_u/k_B = 197$ K) is detected with a good signal-to-noise ratio. Towards the R2 position, the H$_2$O ($3_{03} - 2_{12}$) line and the *para* H$_2$O ($3_{13} - 2_{02}$) line at 2164 GHz ($E_u/k_B = 205$ K) are detected with a good signal-to-noise. The *ortho* H$_2$O ($2_{21} - 1_{10}$) line at 2774 GHz ($E_u/k_B = 194$ K) and the *para* H$_2$O ($4_{04} - 3_{13}$) line at 2392 GHz ($E_u/k_B = 319$ K) are not detected towards B1 and R2, however, in the spaxel centred on (+4.6″,+1.5″), emission is picked up from the central source.

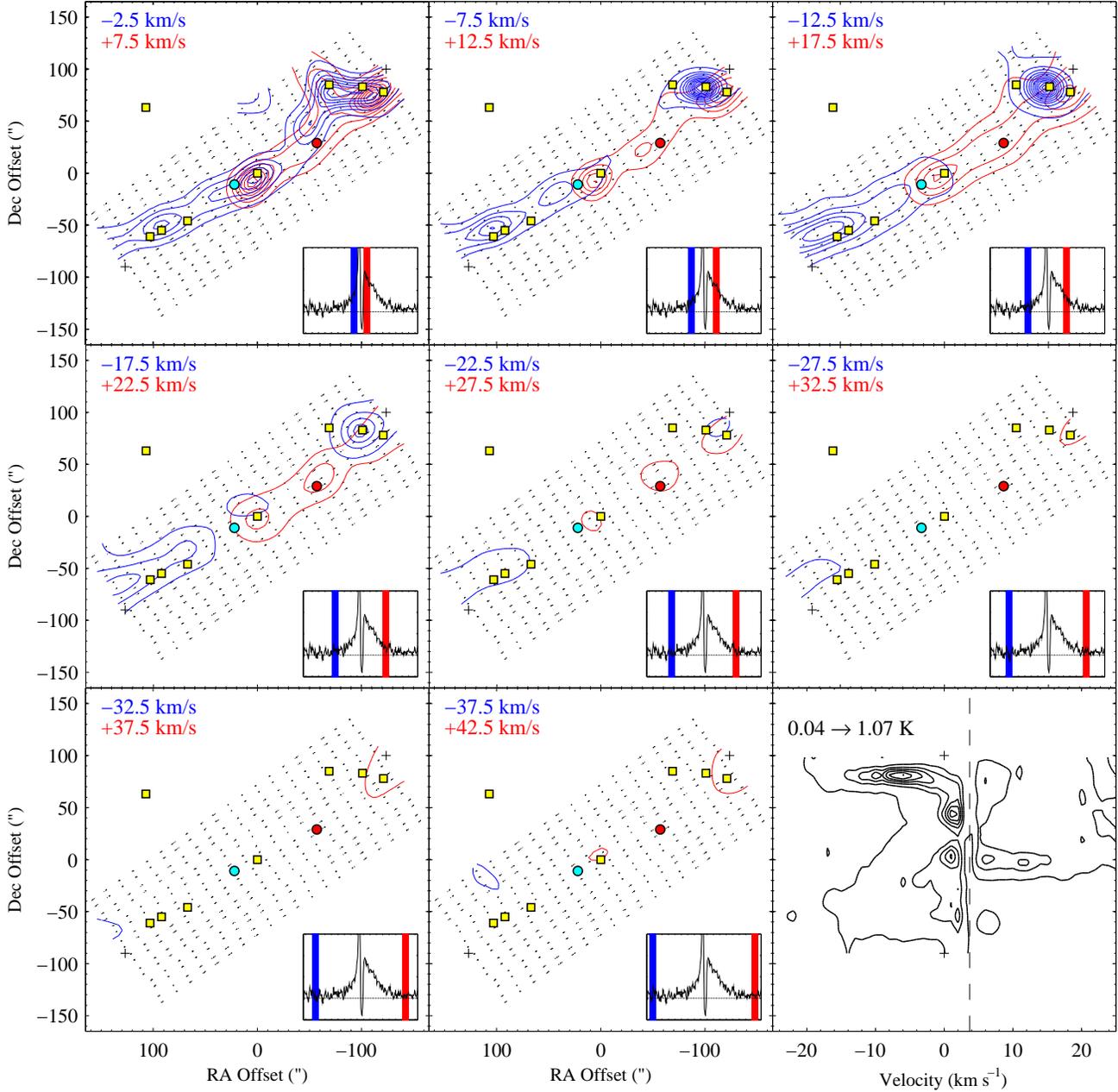

**Fig. 4.** Integrated H$_2$O ($1_{10} - 1_{01}$) intensities in different velocity intervals where $\Delta v = 5$ km s$^{-1}$. The center velocity of the bins is indicated in each panel. Contours are in steps of 0.15 K km s$^{-1}$ and the first contour is at 0.1 K km s$^{-1}$ for all maps except the two maps at lowest velocity, where the first contour is at 0.3 K km s$^{-1}$ and the contour interval is 0.3 K km s$^{-1}$. High velocity gas ($v_{LSR} > 35$ km s$^{-1}$) is observed both close to the central source and further out in the flow. The readout positions for the OTF map are indicated with black dots. The squares and circles represent the positions of the sources presented in Fig. 2. The insets in the lower right corners illustrate each velocity interval for the spectrum towards VLA 1623. The lower right panel shows the position-velocity map in the direction connecting the two plus signs. Here, $v_{LSR} = 3.7$ km s$^{-1}$ is indicated with a dashed line.

## 4. Discussion

### 4.1. Density, temperature, and H$_2$O abundance

#### 4.1.1. Constraints from H$_2$ mid-IR emission

Liseau & Justtanont (2009, and references therein) estimate the H$_2$ density in the outflow to $\sim 3 \times 10^5$ cm$^{-3}$, the tem-perature to $\sim 10^3$ K, and the average column density to $N(\text{H}_2) \approx 3.5 \times 10^{19}$ cm$^{-2}$. These authors used ISO-CAM to observe the 0–0 S(2) to 0–0 S(7) pure rotational lines of H$_2$. We can however not exclude the possibility that the water emission stems from a region with different excitation conditions. For that reason, we have used *Spitzer* H$_2$ line data (Neufeld et al. 2009, Neufeld, private communication) to generate temperature and column density maps of the north-western outflow. From the observed 0–0 S(0) to 0–0 S(7) lines rotational diagrams are made in each pixel of the observed map. These fluxes are corrected for extinction, using the Rieke & Lebofsky (1985) curve and the dust opacities are taken from Ossenkopf & Henning (1994, no ice mantles). The value of the visual extinction is taken to be $A_V = 21$ (Liseau & Justtanont 2009), i.e. equal to the mean value presented in that paper. In reality, the extinction varies from 3 to 50 in the region mapped by ISO-CAM. However, the uncertainty attributed to this variation is less than 20% for the derived mean temperatures in the north-western part of the flow (see below). The rotational diagrams show that the lowest energy transitions trace a gas component that is colder than the gas traced by the lines observed with ISO-CAM. To derive the temperature of the low temperature component, we fit a straight line to the first two $p$-H₂ lines, i.e. the S(0) and S(2) lines. For the hot gas, we fit the S(3) to S(7) lines. For the warm gas component (low temperature in the rotational diagram) we find a temperature of ~400 K while the temperature of the hot gas is estimated at ~1000 K. The column density of hot H₂ (~ $10^{19}$ cm⁻²), obtained from the analysis of Spitzer data, is consistent with the findings presented by Liseau & Justtanont (2009). The column density of the warm component discussed here is, however, significantly higher and peaks towards HH 313 A at $N$(H₂) ≃ 3×10²⁰ cm⁻² (see Fig. 7). The spatial extent of the H₂O outflow agrees with the spatial extent of warm H₂ (see Figs. 2 & 7).

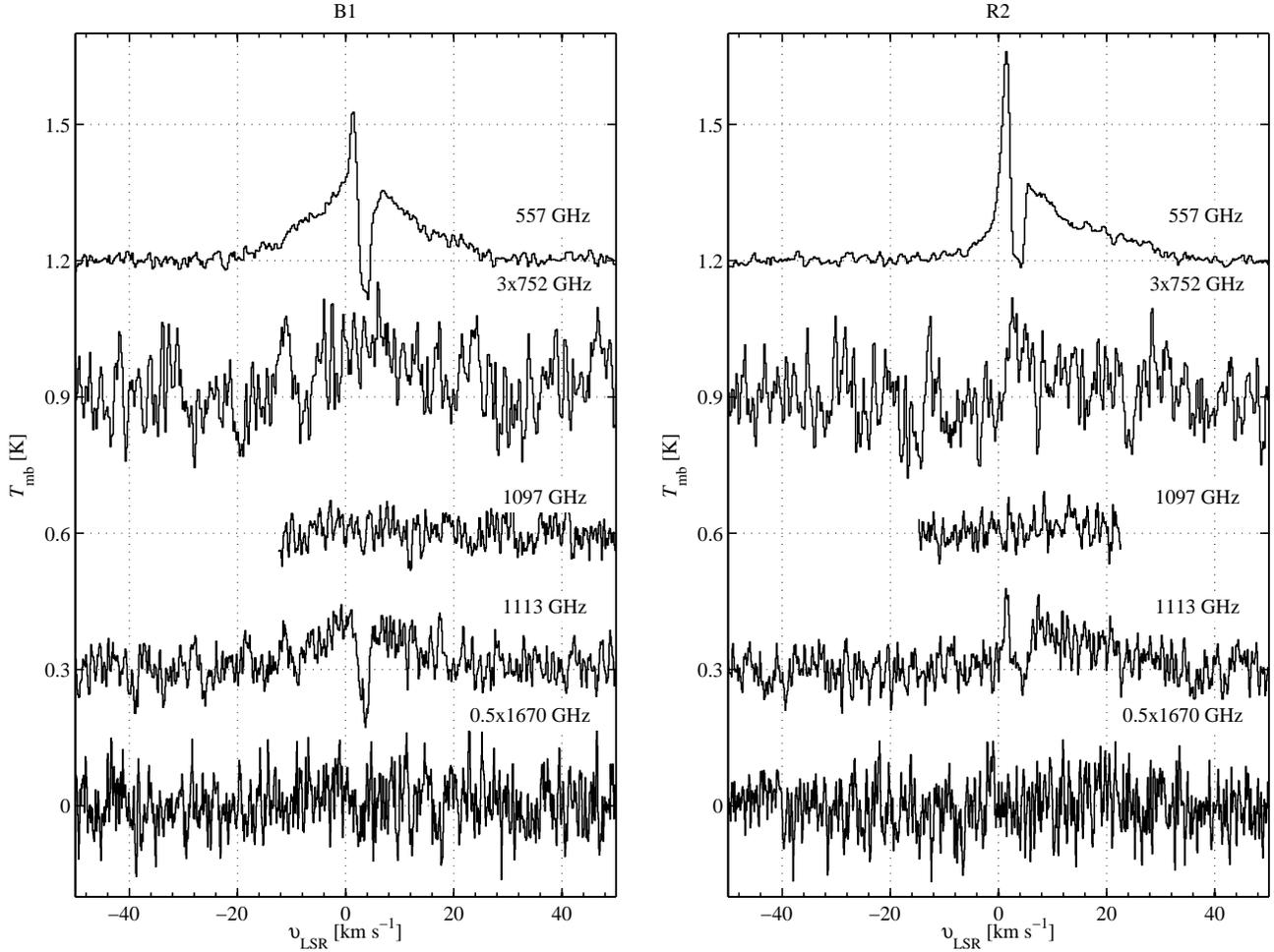

**Fig. 5.** Observed HIFI H$_2$O lines towards the B1 and R2 positions. The lines have been baseline subtracted and are shifted upwards for clarity.

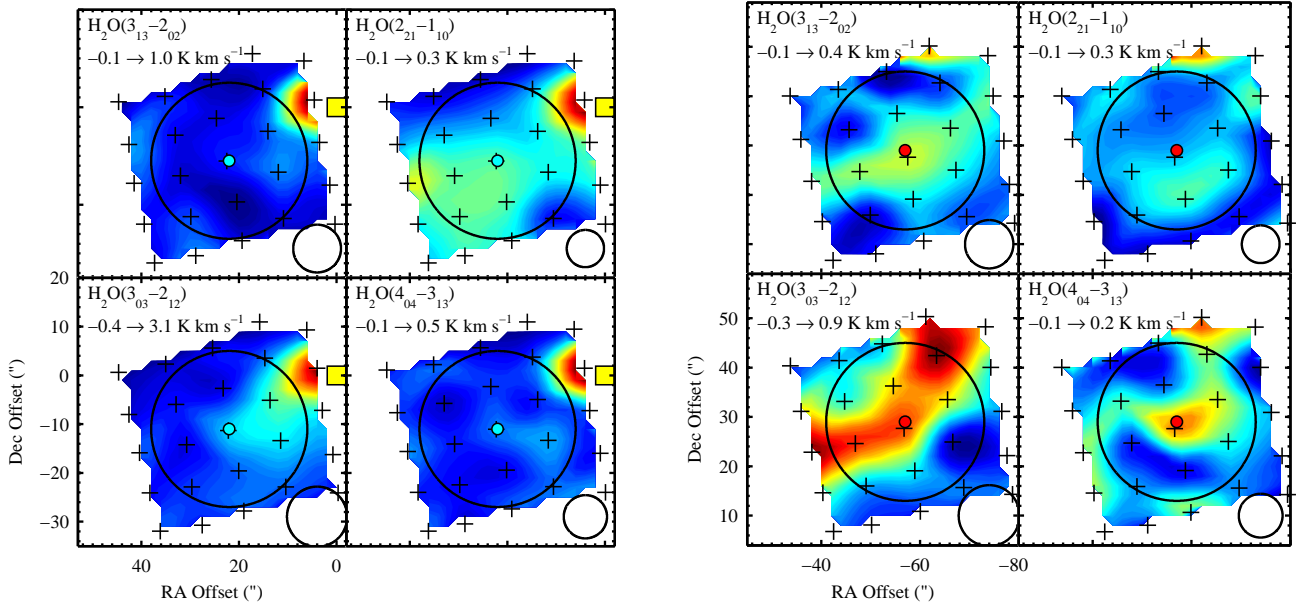

**Fig. 6.** Observed PACS lines towards the B1 (left panel) and R2 (right panel) positions. The blue and red dots indicate the positions of B1 and R2 and the plus signs represent the pointing of each of the 25 spaxels. In the upper left corner of each figure, the line and the range of the colour scale is indicated. Beam sizes are indicated in the lower right corners and the 32″ circular region where the line flux has been averaged is visualised with a black circle. In the left panel, part of the line emission originating in the region close to VLA 1623 (yellow square) is picked up in the nearby spaxel.

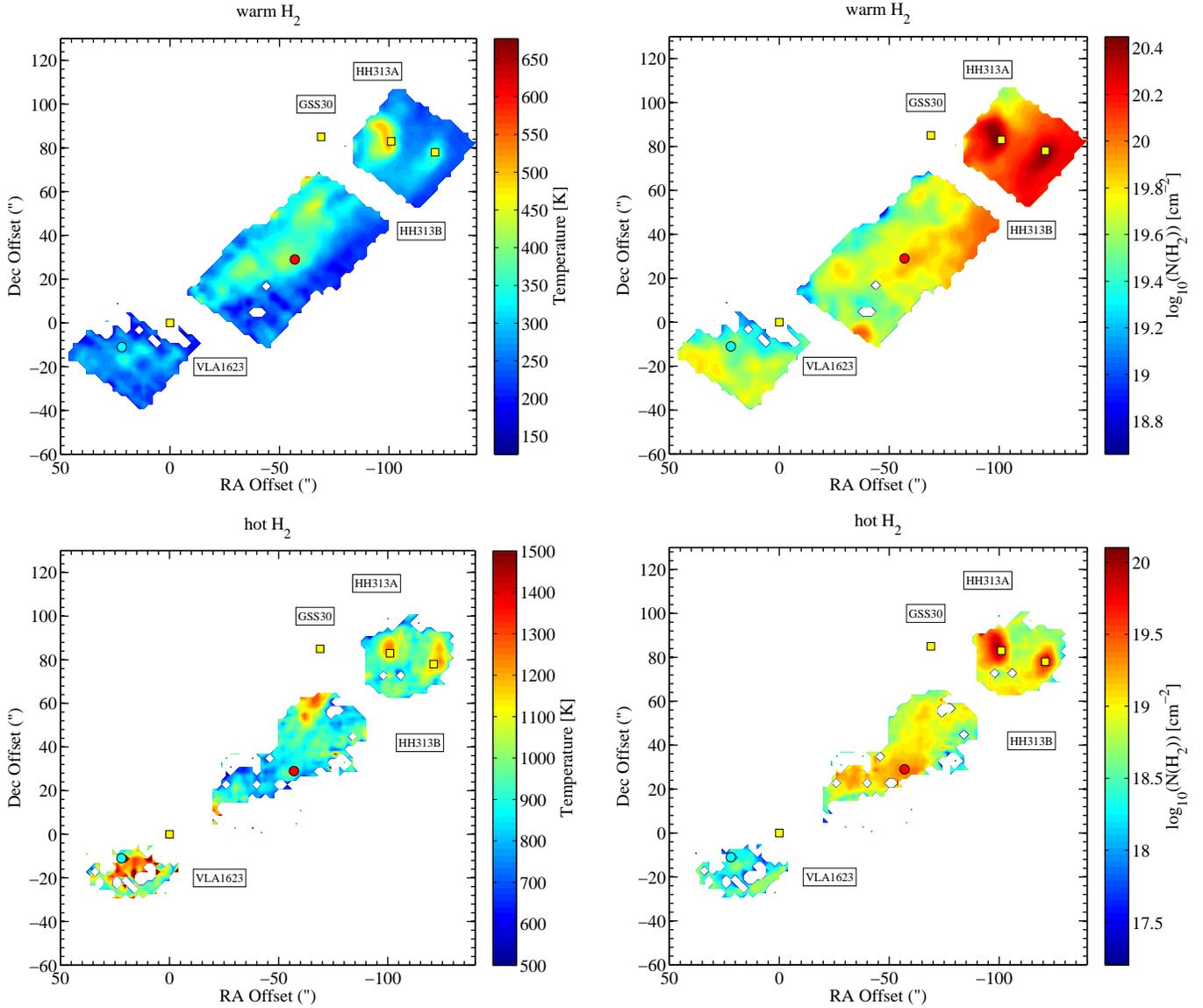

**Fig. 7.** *Left panels:* Temperature of the warm and hot gas, i.e. the low and high temperature components from the H₂ rotational diagrams (see text). *Right panels:* Column density of the warm and hot H₂ gas. In all panels, the positions of R2, B1, VLA 1623, GSS30, HH 313 A and HH 313 B are indicated with coloured dots.

### 4.1.2. H₂O excitation

To interpret the water emission, we have used the non-LTE radiative transfer code RADEX (van der Tak et al. 2007). A grid of models was used, where the temperature was varied from 100 K to 2000 K, the density from $1 \times 10^4$ cm⁻³ to $1 \times 10^8$ cm⁻³ and the water column density from $3 \times 10^{11}$ cm⁻² to $1 \times 10^{15}$ cm⁻². The code was run in LVG (Large Velocity Gradient) mode and the collisional rate coefficients were taken from Daniel et al. (2011). Using these collisional rate coefficients, the estimated H₂O column densities are lower by less than a factor of two, compared to when the collisional rate coefficients presented by Faure et al. (2007) are used.

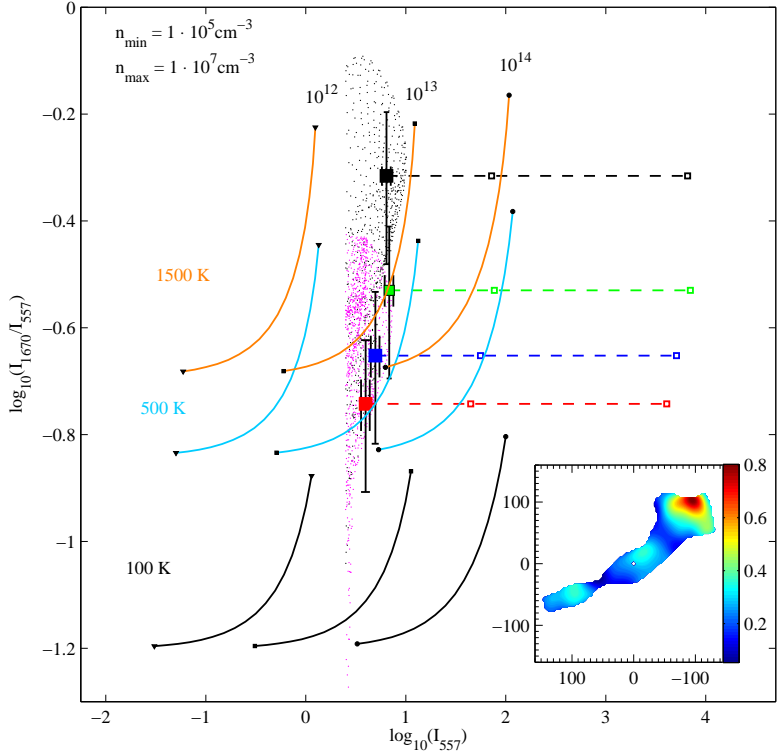

**Fig. 8.** Integrated line strengths as a function of density, temperature and $o$-H₂O column density. The resolution of the PACS (1670 GHz) data has been degraded to match the resolution of the HIFI (557 GHz) data. The black, cyan and orange curves show the line ratio as a function of increasing $o$-H₂O column density (from $10^{12}$ to $10^{14}$ cm⁻²) for three different values on the kinetic temperature of the gas (100, 500, and 1500 K). The lowest column density is to the left in this figure. Within each curve the density of the gas varies from $10^5$ cm⁻³ to $10^7$ cm⁻³ (from left to right). The observed beam averaged values are plotted with coloured squares. Red is towards the position of R2, blue is towards B1, black is towards GSS 30, and green is towards VLA 1623. The dashed lines represent the same regions for smaller source sizes where consecutive squares indicate source sizes of 10″, and 1″. Black and purple dots represent observed ratios towards the region close to GSS 30 and the outflow respectively, i.e. where $\int T_{mb}(557\,\text{GHz})\,d\upsilon \geq 2.5$ K km s⁻¹. The inset in the lower right corner shows the line ratio over the map on a linear scale. Here, the offsets are with respect to VLA1623.

In Fig. 8, we plot the integrated line intensity ratios for the H₂O ($1_{10} - 1_{01}$) and H₂O ($2_{12} - 1_{01}$) transitions versus the integrated line intensity for the H₂O ($1_{10} - 1_{01}$) transition, assuming a constant line width of 20 km s⁻¹. For these models, the optical depths are low (τ ≪ 1) except in the case where the column density is $N(o$-H₂O$) = 10^{14}$ cm⁻². In this case, the optical depths in the two lines are close to unity. When comparing with the observations, a common resolution of 32″ was used for the 557 GHz and 1670 GHz data. We find this to be the optimum resolution since, the B1 position is located very close to the central source and the resolution of the PACS map needs to be degraded in order to obtain clear detections toward the selected outflow positions.

To compare with the observational data we plot the observed total integrated intensity ratios in a few selected positions in the outflow (see Fig 8). If the beam-filling factor is smaller than unity, we assume that the source structure and the beam shape are Gaussian (dashed lines in Fig. 8). We note that the total integrated intensity over the 557 GHz line (including absorption and emission) is higher than the corresponding value measured over the line wings (excluding absorption and emission) by ~40% in the R2 position and by ~20% in the B1 position.

### 4.1.3. VLA 1623 and GSS 30

From Fig. 8 we estimate the beam averaged column density towards the positions of VLA 1623 and GSS 30 to

$N(o$-H₂O$) \gtrsim 1 \times 10^{13}$ cm⁻³. Assuming that the emission is extended over the HIFI beam (coloured squares in Fig. 8), the data are consistent with gas of intermediate to high temperature (i.e. $T > 200$ K). We note that the PACS H₂O ($2_{12} - 1_{01}$) map does not reveal any extended emission towards the position of GSS 30, i.e. the source seems point-like to *Herschel*. The dashed line represents the intensity ratio for a decreasing beam filling factor, down to a source size of 1″. In the case where the beam-filling factor is small, the data are more consistent with a scenario where the column density is higher than $1 \times 10^{13}$ cm⁻².

As mentioned in Sec. 3, enhanced H₂O ($2_{12} - 1_{01}$) emission is detected close to the position of VLA 1623. Comparison with 7 mm interferometer maps from Ward-Thompson et al. (2011) shows that the water emission originates in a region that is extended in the direction orthogonal to the outflow axis but also along the outflow axis (see Fig. 9). The angular extension from the position of VLA 1623 is ~8″ to the north-east and ~5″ to the south-west. These numbers correspond to $9 \times 10^2$ AU and $6 \times 10^2$ AU respectively. In other words, this feature is larger than the measured Full Width Half Maximum (FWHM), 70 AU (53 AU at 120 pc), of the inner disk (Pudritz et al. 1996). The line intensity ratio of the 1670 GHz and the 557 GHz lines allows us to set some limits on the physical parameters. The data are only consistent with a scenario where the temperature is $T > 300$ K and the water column density is higher than $N(\text{H}_2\text{O}) = 1 \times 10^{14}$ cm⁻². Assuming that this emission originates in an extended ellipsoidal structure, with a size comparable to the angular extension, the total water mass has to be at least $10^{23}$ g ($10^{-5}$ $M_\oplus$).

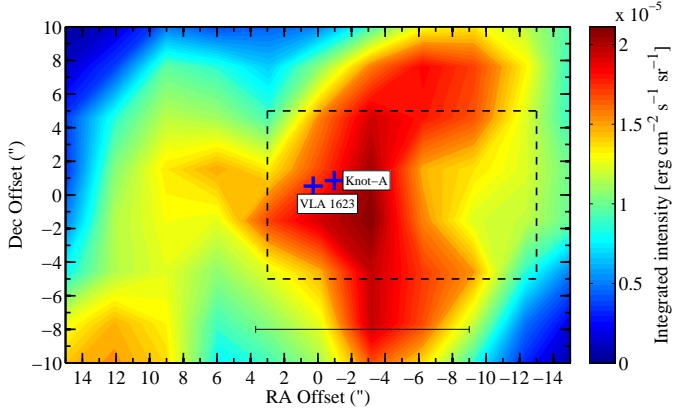

**Fig. 9.** Colour map showing the $H_2O$ $(2_{12} - 1_{01})$ emission observed with PACS. The dashed region indicates the coverage of the SMA, 1.3 mm continuum map presented in Maury & André (2012, their Figure 1). The position of the Class 0 source and Knot-A, discussed in that paper, are indicated with blue crosses. The horizontal bar indicates the beam size of *Herschel* at 1670 GHz.

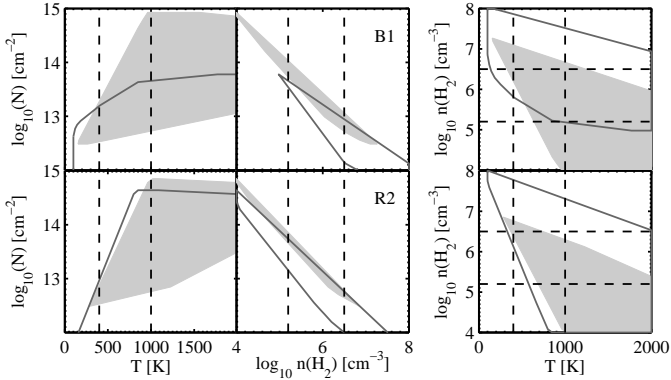

**Fig. 10.** $N(H_2O)$, $n(H_2)$ and $T$, that are in agreement with the observed values within the uncertainties. The upper panels are for B1 and the lower panels for R2. The filled areas are for the *ortho*-lines and the regions indicated with a solid line are for the *para*-lines. The temperature for the warm and hot component and the density range inferred from the [O I] 63 μm to [O I] 145 μm ratio (Liseau & Justtanont 2009) are indicated with dashed lines.

### 4.1.4. R2, B1 and the north-west jet

Towards the R2 and B1 positions, the beam averaged water column densities are estimated to be in the range, $N(H_2O) = (0.03 - 10) \times 10^{14}$ cm⁻². The temperature is estimated to be $T \geq 200$ K at B1 and $T \geq 400$ K at R2. In Fig. 10 we plot the values for $N(H_2O)$, $n(H_2)$ and $T$, that are in agreement with the observed values within the uncertainties. The ortho-to-para ratios for $H_2O$ are estimated to be $2.0^{+1.0}_{-0.3}$ and $2.0^{+1.1}_{-0.7}$ at R2 and B1 respectively. The mean values correspond to a spin temperature of ∼20 K.

The volume densities in the B1 and R2 positions are consistent with the density inferred from the large [O I] 63 μm to [O I] 145 μm ratio presented by Liseau & Justtanont (2009). Assuming that the $H_2$ column density is equal to the value inferred from the *Spitzer* $H_2$ data (warm component), we estimate the water abundance towards B1 as $X(H_2O) \simeq 10^{-7}$.

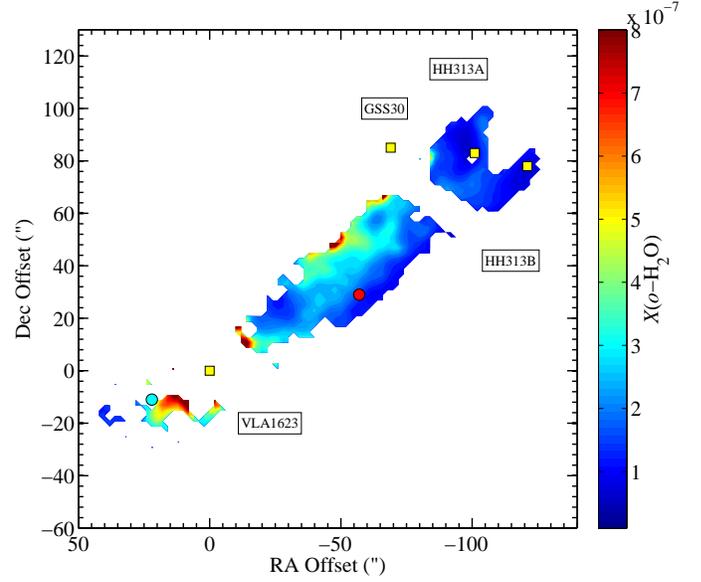

**Fig. 11.** Colour-scale shows the $o$-$H_2O$ abundance variation with respect to $H_2$ in the north-western part of the flow (see Sec. 4.1 & Fig. 7). The positions of R2, B1, VLA 1623, GSS30, HH 313 A and HH 313 B are indicated with coloured dots.

Assuming that the volume density and the kinetic temperature are constant in the north-western flow ($n(H_2) = 10^6$ cm⁻³ and $T = 400$ K ), we estimate the beam averaged column density in all positions of the map from the $H_2O$ $(1_{10} - 1_{01})$ integrated line intensities. Comparison with the warm $H_2$ gas observed with *Spitzer* thus allows to estimate the water abundance in the region (Fig. 11). The water abundance is only presented in the region where the estimated temperature is, $T = 400 \pm 120$ K. Changing the volume density by an order of magnitude will alter the abundance by an equal amount in the opposite direction, due to the fact that the water emitting gas is sub-thermal and at a relatively low density. It should therefore be pointed out that this map merely shows the abundance variation, not the absolute values. The $o$-$H_2O$ abundance is estimated to be a few times $10^{-7}$ with only small variations in the flow. The derived water abundance will be discussed further in Sec. 4.3.

### 4.2. Kinematics and energetics

In this section we estimate the physical parameters of the flow, using the combined information obtained from the $H_2$ and $H_2O$ observations. Assuming that the blue and red flows are reasonably symmetric, these results can be compared with the work presented by André et al. (1990). The characteristic velocity of the $H_2O$ gas can be estimated from the observation of $o$-$H_2O$ at 557 GHz, using the method described in André et al. (1990, their Eq. 5) and averaging all positions in the mapped area. This yields $v_{char} = +16$ km s⁻¹ for the redshifted part of the flow ($v_{LSR} > 8$ km s⁻¹) and $v_{char} = -12$ km s⁻¹ for the blueshifted part of the flow ($v_{LSR} < 0$ km s⁻¹). The column density of warm $H_2$ in the redshifted part of the outflow is typically $N(H_2) \simeq 10^{20}$ cm⁻². The extent of this part of the flow is ∼150″, while the width of the flow is of the order of 20″. Thus the total mass of warm $H_2$ in the red-shifted flow is ∼1.6×10⁻³ $M_\odot$, i.e. slightly lower than what was derived from CO observations (André et al. 1990; Dent et al. 1995). From the flow extent, mass and characteristic velocity, other characteristic flow parameters can be derived. Here we use an inclination correction equal to the one adopted by André et al. (1990), viz. $1/cos(i)$, where $i = 80°$ is the inclination with respect to the line of sight (Method 1). Since the inclination correction is a crucial parameter when deriving the flow parameters, it should be noted that this correction may be too large. Furthermore, the dynamical time scale may be overstimated by an order of magnitude when using the characteristic velocity (see Downes & Cabrit 2007). For that reason we also estimate the parameters of the H$_2$O and H$_2$ gas using the inclination corrections suggested by these authors (Method 2). In particular we do not apply any inclination correction when computing the momentum of the flow. For the kinetic energy, we apply a correction of 5 (see Downes & Cabrit 2007, Fig. 4 and 5) and we use the most recent estimate of the inclination angle, i.e. 75°. The dynamical time scale is computed from the maximum velocity of the flow, i.e. 30 km s$^{-1}$, deprojected by $cos(i)$. A comparison between the derived parameters from CO (2−1) and H$_2$O is presented in Table 3. Uncertainties in this table are based on the assumption that the inclination angle is correct to within 5°. This yields an uncertainty range, on the characteristic velocity of the red-shifted gas, of +45 km s$^{-1}$ < $v_{char}$ < +90 km s$^{-1}$ (Method 2). When using Method 1, the larger characteristic velocity for H$_2$O compared to CO is predominantly due to the shallower shape of the H$_2$O line profile. The momentum rate is independent of the molecular tracer used, i.e. the force is the same. The same applies to the energy and the mechanical luminosity.

### 4.3. Observed line profile shapes

The HIFI 557 GHz map (see Fig. 1) shows a remarkable variety in line profile shapes. In most positions of the map, however, the line cores have an emission peak that is blue-shifted with respect to the absorption feature. This type of line profiles can be a sign of infall and is also commonly observed towards Class 0 sources (Kristensen et al., A&A in press). In the case of VLA 1623, however, infall profiles are observed over a large field of view and they could for that reason be due to the velocity structure of $\rho$ Oph A (Bergman et al. 2011) instead of being caused by large scale infall. The absorption feature is centred at a velocity, with respect to the LSR, similar to the peak velocity of the C$^{18}$O (3−2) emission (Liseau et al. 2010), i.e. at $v_{LSR} \approx 3.7$ km s$^{-1}$. C$^{18}$O peaks in the dark core region, close to the position where the absorption is strongest in the 557 GHz map. It is thus a possible scenario that wide spread H$_2$O emission is observed in the line center and the profile suffers from most absorption towards the dark core at a velocity, slightly offset. In the velocity interval, 2 to 3 km s$^{-1}$, where the 557 GHz emission peaks, also wide spread C$^{18}$O has been detected (see Fig. 4 of Liseau et al. 2010).

As already discussed in Sec. 3, both blue-shifted and red-shifted gas is observed in the north-western region of the map due to the presence of other flows. However, the south-eastern outflow lobe provides a clean case in the sense that it is free from contamination from other flows. From the observed line profiles in this part of the flow we can draw some conclusion regarding the geometry. The inclination angle of the VLA 1623 outflow has been estimated at ∼75° with respect to the line of sight (Davis et al. 1999, from observations of HH313). In the extreme scenario where the flow is laminar and only shear is present, this would yield maximum velocities of ∼ 140 km s$^{-1}$ (assuming that the inclination angle of the south-eastern flow is comparable to the inclination angle of the north-western flow). This is higher than the maximum velocities of ∼ 60 − 80 km s$^{-1}$ that have been estimated from proper motion studies (Caratti o Garatti et al. 2006; Davis et al. 1999). One could argue however, that higher velocities than this are present in the jet of the flow. However, from Fig. 4, there is no obvious indication that molecular gas moving at a high velocity is highly collimated. We therefore conclude that transverse gas motions may play an important role. Hence, the scenario can be compared to the one presented in Bjerkeli et al. (2011) for HH 54 (apart from the fact that the VLA 1623 outflow lies almost in the plane of the sky). These authors conclude that the geometry may be important when it comes to the observed line profile shapes.

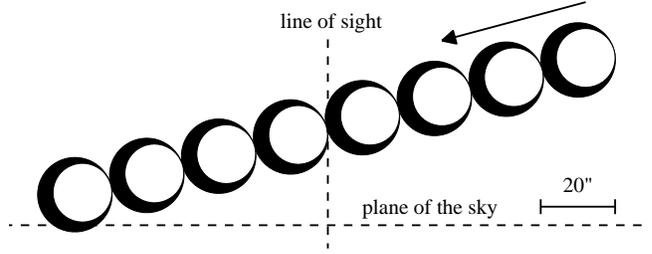

**Fig. 12.** Cartoon image illustrating a cut through the model described in Sec. 4.3. The flow is inclined 75° with respect to the line of sight. In this figure, the thickness of each shell has been increased by a factor of 20 for clarity. The arrow points in the outflow direction and the dashed lines indicate the line of sight and the plane of the sky. Note that the surrounding cloud is not included in this image.

In Sec. 4.1 we put constraints on the H$_2$ density, kinetic temperature, and $o$-H$_2$O column density. To test whether the observed line profiles can be explained by a curved geometry, or not, we use a Monte Carlo code (Hartstein & Liseau 1998) to compute the line profiles, where the most recent collisional rate coefficients have been used (Daniel et al. 2011). The model contains eight spherical shells (see Fig. 12) of a diameter equal to the estimated width of the flow, i.e. $4 \times 10^{16}$ cm (20″ at $d = 120$ pc). The thickness of each shell (see Table 4) attains the highest value in the center of the flow and goes to zero at the edges. The velocity of the gas is varied linearly from 40 to 0 km s$^{-1}$, and the volume density and kinetic temperature is set to $n(H_2) = 1 \times 10^6$ cm$^{-3}$ and $T = 400$ K respectively. The shells are surrounded by a low-density, low-temperature cloud with a diameter of 0.3 pc, responsible for the absorption at $v_{LSR} = 3.7$ km s$^{-1}$.

A synthesized spectrum is presented in Fig. 13, together with the observed spectrum towards the B1 position. In the presented model, the H$_2$O abundance is slightly lower than what was inferred from the observations towards the north-western part of the outflow (see Sec. 4.1 and Fig. 11), i.e. $X(H_2O) = 6 \times 10^{-8}$. Other parameters of the model are presented in Table 4. Due to the relatively large velocity gradients and the curved geometry, the optical depths in the line wing are low ($\tau \ll 1$), although the optical depth at the line center is as high as 11. Gas motions perpendicular to the flow axis, therefore, show up as a red-shifted emission feature. Further out in the south-eastern flow, however, the red-shifted component is less prominent. The spectrum towards (+76″,−38″) is thus not easily reconcilable with a scenario where the inclination angle with respect to the line of sight is as high as 75°. In this position, the line profile appears triangular with no, or little red-shifted emission. The synthesized line profile is in better agreement with the observed spectra in the north-western part of the flow. It should be noted, however, that gas originating from GSS 30 (blue-shifted) is also present in this region.

Previously published H$_2$ (2.12 $\mu$m) images reveal a bow-shaped structure (Davis et al. 1999) towards the position of

**Table 3.** Comparison between the flow parameters derived from CO (2−1) and H₂O. The uncertainty on the inclination is assumed to be 5° and the velocity of the wind is set to 200 km s⁻¹.

| Parameter[c] | Symbol and unit | Method 1[a] ($i = 80°$) | | Method 2[b] ($i = 75°$) |
|---|---|---|---|---|
| | | CO (2−1) | H₂O (red) | H₂O (red) |
| Length to width ratio | $R_{coll}$ | 14 | 7.5 | 7.5 |
| Total mass | $M_{tot}$ ($M_\odot$) | $10 \times 10^{-3}$ | $1.6 \times 10^{-3}$ | $1.6 \times 10^{-3}$ |
| Extent | $R_{max}$ (pc) | 0.08 | 0.09 | 0.09 |
| Characteristic velocity | $v_{char}$ (km s⁻¹) | $50^{+25}_{-25}$ | $90^{+90}_{-30}$ | $60^{+30}_{-15}$ |
| Dynamical time scale | $t_d$ (yr) | $1.7^{+1.6}_{-0.6} \times 10^3$ | $1.0^{+0.5}_{-0.5} \times 10^3$ | $7.4^{+2.4}_{-2.4} \times 10^2$ |
| Momentum | $P$ ($M_\odot$ km s⁻¹) | $0.6^{+0.2}_{-0.3}$ | $0.1^{+0.1}_{-0.05}$ | 0.03 |
| Kinetic energy | $E$ (erg) | $3^{+3}_{-2} \times 10^{44}$ | $1^{+4}_{-0.4} \times 10^{44}$ | $2.0 \times 10^{43}$ |
| Momentum rate | $\dot{P}$ ($M_\odot$ km s⁻¹ yr⁻¹) | $3.4^{+4.2}_{-2.6} \times 10^{-4}$ | $1.5^{4.5}_{-0.8} \times 10^{-4}$ | $3.4^{1.7}_{-0.8} \times 10^{-5}$ |
| Mechanical luminosity | $\dot{E}$ ($L_\odot$) | $1.4^{+3.4}_{-1.1}$ | $1.1^{+7.8}_{-0.8}$ | $0.2^{+0.1}_{-0.04}$ |
| Wind mass-loss rate | $\dot{M}$ ($M_\odot$ yr⁻¹) | $1.7^{+2.1}_{-1.3} \times 10^{-6}$ | $0.7^{+2.3}_{-0.4} \times 10^{-6}$ | $1.7^{+0.8}_{-0.4} \times 10^{-7}$ |

**Notes.** [a] The parameters derived from CO (2−1) were presented in André et al. (1990) and have here been scaled to the distance 120 pc. [b] The inclination corrections are based on the work presented in Downes & Cabrit (2007). The dynamical time-scale is derived from the maximum velocity of the H₂O gas, i.e. ~30km s⁻¹. [c] The length to width ratio and total mass are inferred from the *Spitzer* H₂ observations.

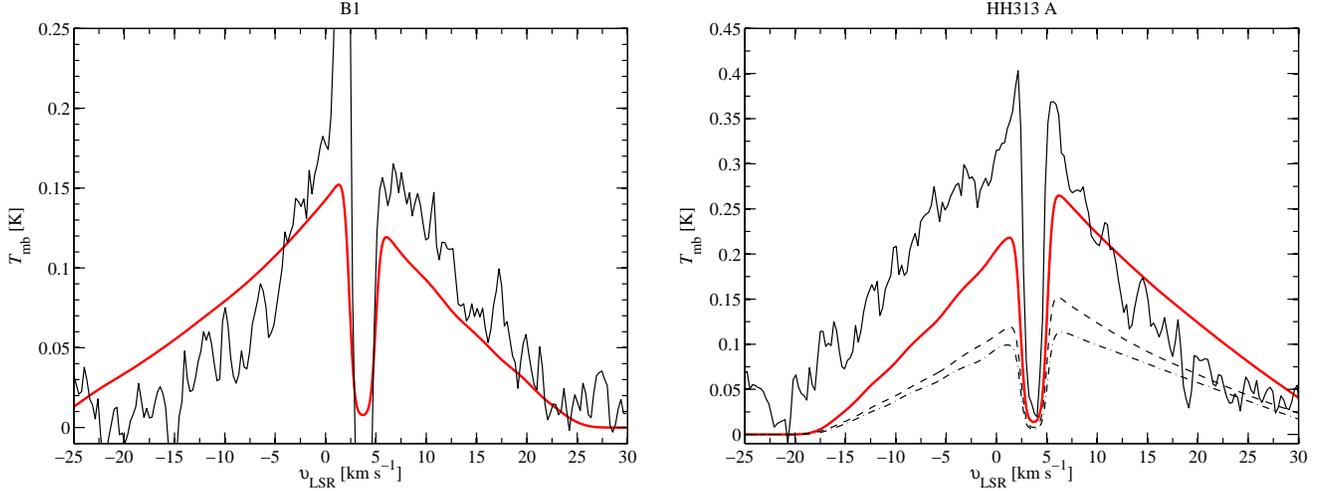

**Fig. 13.** *Left panel:* Observed 557 GHz spectrum (solid black line) at the B1 position compared to the spectrum computed with the Monte Carlo code (solid red line) where $X(H_2O) = 6 \times 10^{-8}$. Note that the narrow emission component is not included in the model. *Right panel:* The observed spectrum (black solid line) at the HH 313 A position compared to the computed spectrum (red solid line). The model has two components, an extended component (dashed line) and a small hot component (dashed-dotted line). Note that blue-shifted emission, originating from GSS 30, is visible in the observed line profile.

**Table 4.** Parameters used in the Monte Carlo models.

| Parameter and symbol | B1 | HH 313 A (hot) |
|---|---|---|
| Distance to source, $d$ | 120 pc | 120 pc |
| H₂O abundance, $X(H_2O)$ | $6 \times 10^{-8}$ | $1 \times 10^{-5}$ |
| LSR velocity, $v_{LSR}$ | + 3.7 km s⁻¹ | + 3.7 km s⁻¹ |
| Velocity profile, $v(r)$ | $40 r$ ($0 < r < 1$) | $40 r$ ($0 < r < 1$) |
| Maximum shell thickness ratio, $\Delta r / R_{max}$ | $5 \times 10^{-2}$ | $1 \times 10^{-2}$ |
| Source size, $\theta_{source}$ | 20″ | 5″ |
| Microturbulence, $v_{turb}$ | 1.0 km s⁻¹ | 1.0 km s⁻¹ |
| H₂ density, $n(H_2)$ | $1 \times 10^6$ cm⁻³ | $1 \times 10^6$ cm⁻³ |
| Kinetic gas temperature, $T_{kin}$ | 400 K | 1000 K |

HH 313 A. This is also the position where the water wing emission is the strongest and the feature is clearly visible in the *Spitzer* H₂ data (see Fig. 7). In this region, hot gas is present at a temperature of ~1000 K, which is why an additional high-temperature component is required to reproduce the observed emission line profile correctly. The red-shifted part of the emission line profile can be explained if an additional shell with a size of 5″, is added to our previous model. The temperature of this emission is taken to be equal to the temperature of the hot H₂ gas (i.e 1000 K) and the volume density is again set to $n(H_2) = 1 \times 10^6$ cm⁻³, i.e. in close agreement with the models presented by Davis et al. (1999). The red-shifted part of the observed emission line profile is satisfactorily reproduced when the water abundance in the hot component is $X(H_2O) = 10^{-5}$.

We conclude that the derived water abundance in the extended flow is less than $10^{-6}$. It can, however, be as high as $10^{-5}$ locally in the shocked region, HH 313 A.

## 5. Conclusions

- We have obtained maps of the two ground-state rotational transitions of *ortho*-water, towards the outflow emanating from VLA 1623. In addition to this, selected transitions of *ortho*- and *para*-water have been observed during a line survey conducted towards two positions in the outflow.
- The H$_2$O ($1_{10} - 1_{01}$) and H$_2$O ($2_{12} - 1_{01}$) lines are detected in the outflow with the HIFI and PACS instruments. In addition to this, the H$_2$O ($2_{11} - 2_{02}$), H$_2$O ($3_{12} - 3_{03}$) and H$_2$O ($3_{03} - 2_{12}$) lines were detected towards the positions B1 and R2. Towards R2, also the H$_2$O ($3_{13} - 2_{02}$) line was detected.
- Observational data are in agreement with LVG models where the column density is $N(\mathrm{H_2O}) \gtrsim 10^{13}$ cm$^{-2}$ towards VLA 1623 and GSS 30 and $N(\mathrm{H_2O}) \simeq (0.03 - 10) \times 10^{14}$ cm$^{-2}$ towards B1 and R2. The temperature of the gas responsible for the H$_2$O emission is higher than 200 K in the observed region.
- Emission at 1670 GHz is detected in a spatially extended region close to VLA 1623. These data are only consistent with a scenario where the column density is higher than $N(\mathrm{H_2O}) \simeq 10^{14}$ cm$^{-2}$.
- The inferred water abundance is lower than what is expected from recent shock models. In the north-western outflow it is estimated to be lower than $10^{-6}$. However, in the HH 313 A bow-shock on spatial scales of $5''$, it is estimated to be $10^{-5}$.
- Estimates of the momentum rate, energy and mechanical luminosity, suggest that the force driving the VLA 1623 outflow is invariant with respect to the molecular tracer used.
- Line profile shapes in the south-eastern flow suggest that transverse gas motions play an important role. The inclination with respect to the line of sight is likely smaller than $75°$.

*Acknowledgements.* The Swedish authors appreciate the support from the Swedish National Space Board (SNSB). Italian authors acknowledge the support from ASI through the contract I/005/011/0. We also thank the HIFI and PACS ICC's for excellent support and the WISH internal referee Lars Kristensen for insightful comments and careful reading of the manuscript.

HIFI has been designed and built by a consortium of institutes and university departments from across Europe, Canada and the United States under the leadership of SRON Netherlands Institute for Space Research, Groningen, The Netherlands and with major contributions from Germany, France and the US. Consortium members are: Canada: CSA, U.Waterloo; France: CESR, LAB, LERMA, IRAM; Germany: KOSMA, MPIfR, MPS; Ireland, NUI Maynooth; Italy: ASI, IFSI-INAF, Osservatorio Astrofisico di Arcetri- INAF; Netherlands: SRON, TUD; Poland: CAMK, CBK; Spain: Observatorio Astronomico Nacional (IGN), Centro de Astrobiologia (CSIC-INTA). Sweden: Chalmers University of Technology - MC2, RSS & GARD; Onsala Space Observatory; Swedish National Space Board, Stockholm University - Stockholm Observatory; Switzerland: ETH Zurich, FHNW; USA: Caltech, JPL, NHSC.

PACS has been developed by a consortium of institutes led by MPE (Germany) and including UVIE (Austria); KU Leuven, CSL, IMEC (Belgium); CEA, LAM (France); MPIA (Germany); INAF-IFSI/OAA/OAP/OAT, LENS, SISSA (Italy); IAC (Spain). This development has been supported by the funding agencies BMVIT (Austria), ESA-PRODEX (Belgium), CEA/CNES (France), DLR (Germany), ASI/INAF (Italy), and CICYT/MCYT (Spain).